\documentclass{aa}  

\usepackage{graphicx}
\usepackage{txfonts}
\usepackage{graphicx}
\usepackage{float}
\usepackage{amsmath}
\usepackage{amssymb}
\usepackage{subcaption}
\usepackage{wrapfig}
\usepackage{xcolor}
\usepackage{booktabs}
\usepackage{color, colortbl}
\usepackage{enumitem}
\usepackage[labelfont=bf]{caption}
\usepackage{forest}
\usepackage[toc,page]{appendix}
\usepackage{chngcntr}

\usepackage[switch]{lineno} 
\linenumbers

\usepackage{filecontents}

\begin{document} 
\nolinenumbers

   \title{A Package for the Automated Classification of Images Containing Supernova Light Echoes}
   \titlerunning{Automated Classification of Images Containing Supernova Light Echoes}

   \author{A. Bhullar
          \inst{1},
          R. A. Ali,
          \inst{2}
          \and
          D. L. Welch \inst{3}
          }

   \institute{University of Guelph,
              Department of Mathematics and Statistics\\
              \email{bhullara@uoguelph.ca}
         \and
             University of Guelph, Department of Mathematics and Statistics\\
             \email{aali@uoguelph.ca}
        \and 
             McMaster University, Department of Physics and Astronomy\\
             \email{welch@physics.mcmaster.ca}
             }

   \date{Received July 15, 2021; accepted August 16, 2021}

 
  \abstract
   {The so-called "light echoes" of supernovae - the apparent motion of outburst-illuminated interstellar dust - can be detected in astronomical difference images; however, light echoes are extremely rare which makes manual detection an arduous task. Surveys for centuries-old supernova light echoes can involve hundreds of pointings of wide-field imagers wherein the subimages from each CCD amplifier require examination.}
   {We introduce ALED, a Python package that implements (i) a capsule network trained to automatically identify images with a high probability of containing at least one supernova light echo, and (ii) routing path visualization to localize light echoes and/or light echo-like features in the identified images.}
   {We compare the performance of the capsule network implemented in ALED (ALED-m) to several capsule and convolutional neural networks of different architectures. We also apply ALED to a large catalogue of astronomical difference images and manually inspect candidate light echo images for human verification.}
   {ALED-m, was found to achieve 90\% classification accuracy on the test set, and to precisely localize the identified light echoes via routing path visualization. From a set of 13,000+ astronomical images, ALED identified a set of light echoes that had been overlooked in manual classification. ALED is available via \protect\url{github.com/LightEchoDetection/ALED}.}
   {}


   \maketitle
%

\section{Introduction} \label{section:paper2_intro}

Supernovae are extremely luminous but transient events that signal the final stage of a massive star's evolution or the disruption of a white dwarf in a close binary \citep{2015ASPC..491..247R}. The bright light from the outburst is radiated in all directions and can be scattered by interstellar dust as outburst light encounters such material. The deflection of light off interstellar dust is analogous to the deflection of sound waves off surfaces; hence the term "light echo". Light echoes from a historical supernova can arrive at Earth centuries later than any direct light and consequently they can facilitate the study of historic supernovae in modern times with modern instrumentation. The light echoes of supernovae are present in some astronomical images, but are almost without exception a small percentage of the surface brightness of the moonless night sky \citep{mcdonald}.

\begin{figure}
    \centering
    \includegraphics[width=.45\textwidth]{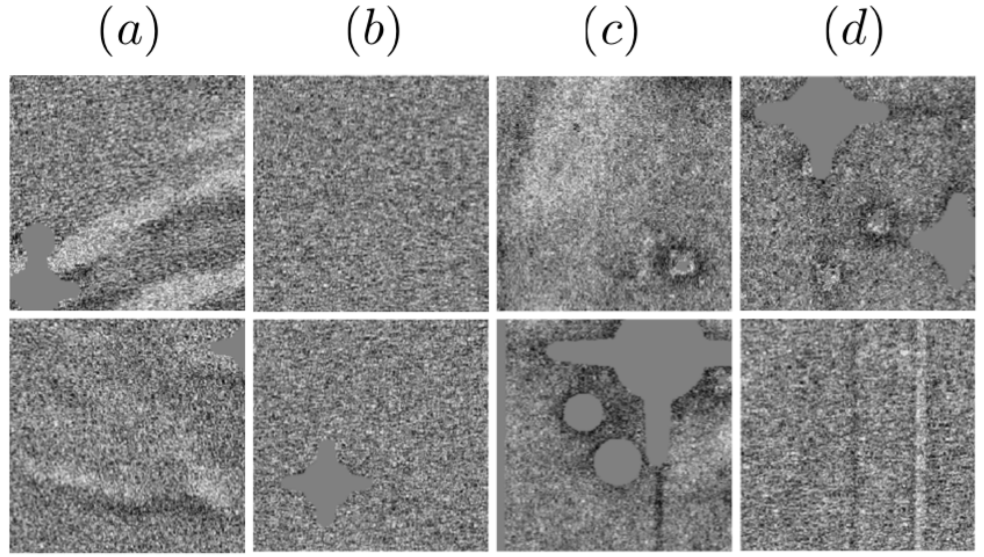}
    \caption{Types of entities present in the CFHT dataset. (a) images that clearly contain at least one light echo; (b) images that clearly do not contain at least one light echo; (c) images that contain light echoes and artifacts; and (d) images that contain entities with light echo-like characteristics. Illustration and caption credit: \citep{rpv}.}
    \label{fig:cfht2}
\end{figure}

A difference image is the result of subtracting a pair of images that are taken at the same telescopic pointing but at different dates. Difference imaging allows for objects that appear to move relative to the background of space, such as light echoes, to be visually detected more easily (see Figure \ref{fig:cfht2}). However, manual visual image analysis is still demanding due to the large amount of data generated by supernova light echo surveys. For instance, in \citet{mcdonald} 13,000+ difference images were individually inspected for light echoes over the course of a year.

This paper introduces the Python package ALED (pronounced "A-led") for automated light echo detection in astronomical difference images. This package provides an invaluable resource for astronomers requiring rapid identification of images containing at least one light echo, and where those light echoes are located within the identified images. ALED takes as input a grey-scale difference image of arbitrary size, and outputs a corresponding routing path visualization of the input image \citep{rpv}. The routing path visualization reveals the regions of the input image that have a high probability of containing a light echo - the brighter the region the greater the likelihood. Here, we demonstrate and compare the performance of ALED-m, the capsule network model that is available in ALED, to several different artificial neural network classifiers.  We also apply ALED to the 13000+ difference images for which \citet{mcdonald} did not detect light echoes. ALED was developed specifically for the purpose of light echo detection. Difference imaging also generates artifacts that can, in some instances, mimic the appearance of light echoes, see Figure \ref{fig:cfht2}. Simpler automatic image classification techniques struggle to distinguish light echoes from light-echo-like artifacts \citep{mcdonald}.

ALED is based on a capsule network \citep{sabour2017dynamic}, which is an extension of a convolutional neural network (CNN). CNNs have achieved state-of-the-art performance on many computer vision problems, and are currently one of the most popular algorithms for image classification. CNN architectures typically contain millions of weights that are to be learned during training, and as a result, require a large training set to prevent over-fitting \citep{lecun1989backpropagation}. For example, LeNet, a CNN containing 5,978,677 weights, was trained using 60,000 images to classify hand-written digits \citep{lecun1998gradient}. This particular classification task is considered simpler than light echo detection from difference images because each image is guaranteed to contain a single white digit in the center of a black background. More complex classification tasks require CNNs with a higher weight count. For example, Wide ResNet-50-2, a CNN containing 68,951,464 weights, was trained using 1.2 million images to classify photographs of 1000 object categories such as boats, horses, etc. \citep{zagoruyko2016wide}. This is a more complex classification task because an image can contain 1 of 1000 coloured objects in an arbitrary orientation on an arbitrary background. 

The difficulty of light echo classification falls somewhere in between these two examples. Although there are only two classification categories - light echo present or not - light echoes are not only found in widely different orientations and backgrounds in grey-scale images but each image may contain several entities, and of those entities, more than one may be a light echo. As such, it is expected that a CNN with millions of weights would be required for adequate light echo detection. In specialized fields, such as observational astronomy, sufficiently large sets of labeled data are often unavailable. This is especially true for automating the classification of images containing supernova light echoes because detectable light echo examples are so few in number \citep{mcdonald}. 

Training a CNN, which contains millions of weights, by using a dataset of a few hundred images is problematic because the model may over-fit to the small dataset in a manner that is not obvious. For example, if the training set represents the validation and test sets well then the validation and test accuracies will be similar to the training accuracy. However, if the model over-fits to irrelevant characteristics of the dataset, such as camera or pre-processing characteristics or artifacts, then the training, validation, and test accuracies will still be similar even though the model will not perform well on datasets taken by different cameras or pre-processed in a slightly different manner.

Capsule network models require far fewer weights, and thus, are less likely to over-fit to irrelevant characteristics of the dataset. Dropout can be used in CNNs to reduce over-fitting by randomly omitting a certain percentage of neurons every forward propagation step while training. This strategy makes it more difficult for the network to simply memorize the training set or learn unnecessary co-adaptations such as higher-level neurons correcting the mistakes of lower-level neurons \citep{dropout}. For completeness, we include a few CNN architectures to illustrate how many more weights are needed in a CNN compared to a capsule network for light echo detection.

Capsule networks are artificial neural networks that can model hierarchical relationships \citep{sabour2017dynamic}. They typically contain fewer trainable weights than CNNs and, as a result, require a smaller training set to achieve good performance. Capsule networks also facilitate routing path visualization which can be used to interpret the entity that a given capsule detects. Said differently, capsule networks provide a way to see what is being detected. It should be noted that CNNs can also facilitate several gradient based methods to see what is being detected, such as grad-cam \citep{selvaraju2017grad}.  

Interpreting how a model functions is important because it will allow for the identification of scenarios that may result in model failure. When the training set is small, as seen in light echo detection, it is expected that the training set will not contain such difficult image classification scenarios even though they are likely to arise in practice. Consequently, constraints can be applied to the network, or improvements can be made to the training set to decrease the probability of failure. 

Section \ref{section:paper2_cfht_dataset} describes the Canada-France-Hawaii Telescope (CFHT) difference image data used to train and evaluate ALED. Section \ref{section:paper2_clf} briefly reviews capsule networks and describes the network architectures to which ALED was compared and the parameter settings used to classify the 13000+ CFHT difference images. Section \ref{section:paper2_models} presents the results from the analyses and concluding comments are made in Section \ref{section:paper2_summary}.

\section{CFHT Dataset} \label{section:paper2_cfht_dataset}

In 2011, CFHT's wide-field mosaic image, MegaCam, was used to conduct a survey with the primary objective of discovering supernova light echoes in a region where three historical supernovae were known to have occurred. Out of the 13,000+ $2048 \times 4612$ difference images that were produced from the survey, 22 were found to contain at least one light echo \citep{mcdonald}. The 22 light echo containing CFHT difference images of size $2048 \times 4612$ were reduced to 350 difference images of size $200 \times 200$, among which 175 contained at least a portion of a light echo and the remaining 175 contained other astronomical entities. The dataset was created by manually masking the light echoes present in the 22 images of size $2048 \times 4612$, which were then cropped to size $200 \times 200$. If a cropped image contained at least 2500 pixels of mask then it was classified as containing a light echo.

Among the 4885 $200 \times 200$ cropped images that did not contain a light echo, 175 were selected at random, along with 175 that were classified as containing a light echo to form the final dataset of 350 images. The dataset consists of images and their respective binary labels, where the label indicates whether the image does or does not contain at least one light echo. The dataset was split into a training set of 250 images, and a validation and test set of 50 images each. See Appendix A for a summary of the pre-processing steps required to produce a difference image.

\section{Methods} \label{section:paper2_clf}

A capsule is a vector whose elements represent the instantiation parameters of an entity in the image, where an entity is defined as an object or object part. The instantiation parameters of an entity are defined as the total information that would be required to render the entity. For example, the instantiation parameters of a circle whose center is positioned at coordinates $(x, y)$ could be given by $(x, y)$ and the radius of the circle. A capsule network consists of multiple layers of capsules, where a capsule in a given layer detects a particular entity. Capsules in a layer are children to the capsules of its succeeding layer, and parent to the capsules of its preceding layer. The child-parent coupling of some pairs of capsules is stronger than others, and this is determined by the routing algorithm \citep{sabour2017dynamic}. 

Initially, when an image is fed into a capsule network, the image is convolved with a set of filters to produce a feature map, i.e., to produce a tensor representation of the image that is better suited for classification \citep{goodfellow2016deep}. The size of the feature map is determined by the length, $F$, and stride, $S$, of the filters, and the total number of filters used, $N$. Higher level feature maps can be produced from these feature maps, giving rise to $M$ feature maps in total. The $M^{th}$ feature map is convolved with a set of filters to produce a convolutional capsule (ConvCaps) layer. The ConvCaps layer contains $I$ capsule types each of dimension $D$, where all capsules of a particular type detect the same entity. The elements of a capsule in the ConvCaps layer are calculated using local regions of the image, so ConvCaps capsules detect simpler entities. Succeeding layer capsules are calculated using preceding layer capsules and transformation weight matrices. Hence, higher-level capsules look at broader regions of the image, and are expected to detect more complex entities.  

Each capsule in the final capsule layer is forced to detect a particular entity, such as a light echo, whilst intermediate layer capsules are not specified to which entities they may detect. The Euclidean length of a capsule will be close to 1 if the entity that the capsule detects is present in the image and close to 0 otherwise. This phenomenon is enforced by minimizing the margin loss while training,
\[L = T \max (0, m^{+} - ||\mathbf{v}||)^2 + \lambda(1-T)\max (0, ||\mathbf{v}|| - m^{-})^2, \]
where $T=1$ if the entity is present in the image and 0 otherwise, $m^{+} = 0.9$, $||\mathbf{v}||$ is the length of the capsule, $\lambda = 0.5$, and $m^{-} = 0.1$ \citep{sabour2017dynamic}. The total margin loss is calculated by summing the margin loss of each capsule in the final capsule layer. If the Euclidean length of the capsule is close to 1, it is said to be ``active".

\begin{figure*}
    \centering
    \includegraphics[width=\textwidth]{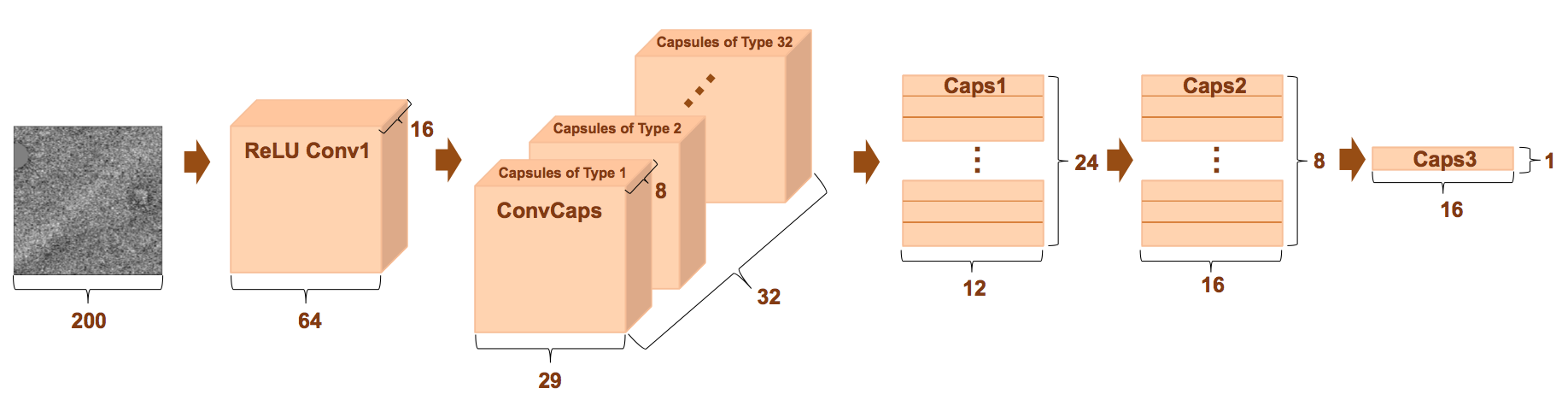}
    \caption{A diagram of the architecture of ALED-m, where ReLU Conv1 is the feature map layer, and Caps1, Caps2, Caps3 are the capsule layers. }
    \label{fig:arch}
\end{figure*}

The architecture of ALED-m is shown in Figure \ref{fig:arch}. To elaborate, a $200\times200$ image is convolved with a set of 16 filters sized $9\times9$ using a stride of 3 to produce the feature map. The feature map is convolved with a set of 256 filters sized $5\times5$ using a stride of 2 to produce the ConvCaps layer. The subsequent capsule layers contain 24, 8, and 1 capsules of dimensions 12, 16, and 16, respectively. The dimensions of higher-level capsules are larger because more degrees of freedom are required to store the instantiation parameters of more complex entities. The length of the capsule in capsule layer 3 will be close to 1 if a light echo is present in the image, and 0 otherwise. ALED also uses weight sharing in which capsules of the same type within a ConvCaps layer share a set of transformation matrices \citep{rpv}.

The entity that a given capsule detects can be localized, if it is present, by routing path visualization \citep{rpv}. Hence, this technique can be used to interpret the entities detected by intermediate layer capsules. In turn, scenarios can be identified that may result in model failure. For instance, there is one capsule in capsule layer 3, and as a result, the information present in all active capsules in the preceding layer, capsule layer 2, will be strongly routed to that single capsule \citep{sabour2017dynamic}. Consequently, if a capsule(s) in capsule layer 2 is detecting an irrelevant object, such as an artifact, than it may adversely influence predictions made by the light-echo-detecting capsule. Since there is only a single capsule in capsule layer 3, the routing path visualization of that capsule will simply be an addition of the routing path visualizations of the capsules in the preceding layer, capsule layer 2. Changes to the training set or to the model can help ensure that all capsules are detecting sensible objects.

\subsection{Model Comparisons}
We compared the performance of ALED-m to ten other capsule network architectures (Table \ref{table:model_caps}), and to five CNN models (Table \ref{table:model_conv}) for baseline reference. ALED-m has a similar architecture to model 8, but contains more filters, capsules, and dimensions per capsule. Models 12 and 13 used the same CNN architecture as model 14, but were trained using augmented training sets of 1000 and 500 images, respectively. The original training set of 250 images was flipped horizontally to create the augmented set of 500 images, and then flipped vertically to create the augmented set of 1000 images in total.

All models were trained using a single NVIDIA Pascal P100 GPU on TensorFlow \citep{abadi2016tensorflow} with the Adam optimizer \citep{kingma2014adam}. The learning rate was initially set to 0.001 because it was found to be the largest learning rate that caused a steady decline in the total margin loss. If the total margin loss of the validation set had not decreased for 10 consecutive iterations then the learning rate was decreased by a factor of 10. Learning was terminated when the learning rate fell below $10^{-6}$. Training was terminated just before the model began over-fitting to the training set. A batch size of 5 was used because it was the largest size that could fit into memory.

An image was classified as containing a light echo if the length of the final capsule was greater than 0.5. The accuracy of the trained model was defined by 
\[ \frac{\texttt{True Positives} + \texttt{True Negatives}}{\texttt{No.} \texttt{ of Images in Set}} \times 100\%,\]
and was found to be $90\%$ on the test set and $88\%$ on the validation set.

\section{Results} \label{section:paper2_models}

Using lower weight CNNs (less than 5 million weights) was necessary to prevent over-fitting to the small training set (less than 500 images). Regardless, due to the large number of weights in the CNNs relative to the small training set, the two fully connected layers before the output layer required a 95\% dropout to combat over-fitting (Table \ref{table:model_conv}). Since dropout is typically set to 50\% \citep{dropout}, it may be that too many neurons were initialized in the fully connected layers thereby resulting in strong over-fitting. Model 15 was trained on the augmented training set of 1000 images to see how a CNN model with less than one million trainable weights, and a more standard percent dropout of 50\% would perform.

\begin{table*}
\centering
\caption{List of capsule network architectures trained on the CFHT dataset. Models are defined by the feature map(s), ConvCaps layer, and capsule layer(s), where $C$ is the number of capsules in the layer, $D$ is the number of dimensions per capsule, $F$ is the length of each filter, $N$ is the number of filters used, $S$ is the stride, $I$ is the number of capsule types, and $M$ is the number of feature maps. The Number F. Maps column lists the total number of feature maps in each model. The Total Weights column lists the total number of trainable weights in each model. All models were trained on 250 images.}
\scalebox{0.86}{
\begin{tabular}{ c c r r r r r r }
\toprule
 & \multicolumn{3}{c}{\textbf{Convolutional Layers}} & \multicolumn{3}{c}{\textbf{Capsule Layers} $(C, D)$} \\
 \cmidrule(lr){2-4} \cmidrule(lr){5-7}
 & \textbf{Number} & \textbf{Feature Maps} & \textbf{ConvCaps} & & & & \textbf{Total}\\
 \textbf{Model} & \textbf{F. Maps} & $M \times (F, N, S)$ & $(F, I, D, S)$ & 1 & 2 & 3 & \textbf{Weights}\\

\midrule
1 & 5 & 3 $\times$ (5, 256, 1)  & (5, 40, 12, 2) & (30, 15) & (\hspace{6pt}1, 25) & & 9,861,010\\
 & & 2 $\times$ (5, 256, 2) & &  \\
\addlinespace
2 & 2 & 2 $\times$ (5, 256, 2) & (5, 40, 12, 2) & (30, 15) & (\hspace{6pt}1, 25) & & 4,945,042 \\
\addlinespace
3 & 2 & 2 $\times$ (5, 256, 2) & (5, 40, 12, 2) & (\hspace{5pt}1, 15) & & & 4,724,992\\
\addlinespace
4 & 1 & 1 $\times$ (9, 256, 3) & (5, 40, 12, 2) & (30, 15) & (10, 25) & (1, 25) & 3,428,222\\
\addlinespace
5 & 1 & 1 $\times$ (9, 256, 3) & (5, 40, 12, 2) & (30, 15) & (\hspace{6pt}1, 25) & & 3,320,722 \\
\addlinespace
ALED-m & 1 & 1 $\times$ (9, \hspace{5pt}16, 3) & (5, 32, \hspace{5pt}8, 2) & (24, 12) & (\hspace{6pt}8, 16) & (1, 16) & 216,608  \\
\addlinespace
6 & 1 & 1 $\times$ (9, \hspace{5pt}16, 3) & (5, 24, \hspace{5pt}8, 2) & (16, 10) & (\hspace{6pt}4, 12) & (1, 12) & 117,280 \\
\addlinespace
7 & 1 & 1 $\times$ (9, \hspace{5pt}16, 3) & (5, \hspace{5pt}6, \hspace{5pt}4, 2) & (\hspace{5pt}8, \hspace{5pt}6) & (\hspace{7pt}4, \hspace{4pt}8) & (1, \hspace{6pt}8) & 13,880 \\
\addlinespace
8 & 1 & 1 $\times$ (9, \hspace{10pt}8, 3) & (5, \hspace{5pt}4, \hspace{5pt}4, 2) & (\hspace{5pt}8, \hspace{5pt}6) & (\hspace{7pt}3, \hspace{4pt}8) & (1, \hspace{6pt}8) & 5,984\\
\addlinespace
9 & 1 & 1 $\times$ (9, \hspace{10pt}8, 3) & (5, \hspace{5pt}2, \hspace{5pt}4, 2) & (\hspace{5pt}6, \hspace{5pt}4) & (\hspace{7pt}2, \hspace{4pt}6) & (1, \hspace{6pt}6) & 2,816 \\
\addlinespace
10 & 1 & 1 $\times$ (9, \hspace{10pt}8, 3) & (5, \hspace{5pt}2, \hspace{5pt}3, 2) & (\hspace{5pt}6, \hspace{5pt}4) & (\hspace{7pt}3, \hspace{4pt}6) & (1, \hspace{6pt}6) & 2,546 \\
 
\bottomrule
\end{tabular}}

\label{table:model_caps}
\end{table*}

\begin{table*}
\centering
\caption{List of CNN architectures trained on the CFHT dataset. Models are defined by the feature maps, number of fully connected layers, and \% dropout per fully connected layer, where $F$ is the length of each filter, $N$ is the number of filters used, $S$ is the stride, and $M$ is the number of feature maps. The Number F. Maps column lists the total number of feature maps in each model. The Total Weights column lists the total number of trainable weights in each model.}
\scalebox{0.82}{
\begin{tabular}{ c c c r c c c c }
\toprule
 &  &  & \multicolumn{3}{c}{\textbf{Fully Connected (FC) Layers}} \\
\cmidrule(lr){4-6}
 & \textbf{Train} &\textbf{Number} & \textbf{Feature Maps} & \textbf{Number} & \textbf{Number} & \textbf{Dropout} & \textbf{Total}\\
\textbf{Model} & \textbf{Size} & \textbf{F. Maps} & $M \times (F, N, S)$ & \textbf{FC Layers} & \textbf{Neurons} & \textbf{\%} & \textbf{Weights} \\
\midrule
11 & 250 & 3 & 2 $\times$ (5, 256, 2) & 3 & 328 & 0  & 22,848,778\\
 & & & 1 $\times$ (5, 128, 2) & & 192 & 50 \\
 & & & & & 2 & - \\
\addlinespace

12 & 1000 & 3 & 1 $\times$ (9, 256, 3) & 3 & 152 & 95 & 4,551,906\\
 & & & 2 $\times$ (5, 128, 2) & & 88 & 95 \\
 & & & & & 2 & - \\
\addlinespace

13 & 500 & 3 & 1 $\times$ (9, 256, 3) & 3 & 152 & 95 & 4,551,906\\
 & & & 2 $\times$ (5, 128, 2) & & 88 & 95 \\
 & & & & & 2 & -  \\
\addlinespace

14 & 250 & 3 & 1 $\times$ (9, 256, 3) & 3 &  152 & 95 & 4,551,906\\
 & & & 2 $\times$ (5, 128, 2) & & 88 & 95  \\
 & & & & & 2 & - \\
 \addlinespace

15 & 1000 & 3 & 1 $\times$ (9, \hspace{6pt}16, 3) & 3 &  152 & 50 & 875,586 \\
 & & & 2 $\times$ (5, \hspace{6pt}32, 2) & & 88 & 50 \\
 & & & & & 2 & - \\
 
\bottomrule
\end{tabular}}
\label{table:model_conv}
\end{table*}

\begin{table*}
\centering
\caption{The prediction accuracy on the validation and test sets is listed for each model. The time taken to train each model is given under the Duration column. The total number of times the weights were updated during training is given under the Steps column. The time taken to classify a single image is given under the Time column, and this number is averaged over 50 images.}
\scalebox{0.9}{
\begin{tabular}{ c c r r r r r r }
\toprule
 &  &  & \multicolumn{2}{c}{\textbf{Accuracy}} & \multicolumn{2}{c}{\textbf{Training}} \\
\cmidrule(lr){4-5} \cmidrule(lr){6-7}
\textbf{Network Type} & \textbf{Model} & \textbf{Weights} & \textbf{Validation} & \textbf{Test} & \textbf{Duration} & \textbf{Steps} & \textbf{Time} (s)\\
\midrule
Capsule & 1 & 9,861,010 & 80\% & 78\% & 14h & 7860 & -\\
& 2 & 4,945,042 & 84\% & 86\% & 15h & 7692 & 0.35\\
& 3 & 4,724,992 & 50\% & 50\% & 2h & 11990 & 0.013\\
& 4 & 3,428,222 & 88\% & 88\% & 11h & 2965 & 0.79\\
& 5 & 3,320,722 & 82\% & 82\% & 11h & 3250 & 0.83\\
& ALED-m & 216,608 & 88\% & 92\% & 4h & 1620 & 0.63\\
& 6 & 117,280 & 88\% & 92\% & 2h & 1950 & 0.22 \\
& 7 & 13,880 & 88\% & 86\% & $<$ 1h & 3960 & 0.022\\
& 8 & 5,984 & 88\% & 90\% & $<$ 1h & 9200 & 0.015\\
& 9 & 2,816 & 84\% & 90\% & $<$ 1h & 2320 & 0.0085\\
& 10 & 2,546 & 86\% & 84\% & $<$ 1h & 2015 & 0.0086\\
\addlinespace

Convolutional & 11 & 22,848,778 & \multicolumn{2}{c}{Over-Fitting} & $<$ 1h & 456 & - \\
& 12 & 4,551,906 & 88\% & 86\% & $<$ 1h & 1653 & 0.0029\\
& 13 & 4,551,906 & 82\% & 86\% & $<$ 1h & 360 & 0.0029\\
& 14 & 4,551,906 & 60\% & 66\% & $<$ 1h & 234 & 0.0029\\
& 15 & 875,586 & 66\% & 60\% & $<$ 1h & 1000 & 0.0016\\

\bottomrule
\end{tabular}}
\label{table:model_accuracy}
\end{table*}

\subsection{Model Evaluations} \label{section:paper2_artifacts}

The accuracy and time taken to classify a single image is summarized in Table \ref{table:model_accuracy}. ALED-m can classify a $200 \times 200$ difference image in 0.63 seconds, and a corresponding routing path visualization can be produced in approximately 2 seconds. For the capsule network models, it was found that classification accuracy and time taken to classify an image increased as the number of capsule layers increased. Increasing the number of feature maps, $M$, and the size of the ConvCaps layer did not noticeably affect the classification accuracy, as demonstrated by models 1, 2, 3. The total number of trainable weights in a model was mainly influenced by the size of the feature map(s) and ConvCaps layer, and not the number of capsule layers. A notable difference between capsule network and CNN models is that capsule networks can be trained to achieve good classification accuracy using a small non-augmented training set, and an architecture that contains a few thousand trainable weights. For instance, capsule network model 8 and CNN model 12 have similar accuracy, but model 8 contains 5,984 trainable weights where as model 12 contains 4,551,906.

\begin{figure}
    \centering
    \includegraphics[width=.45\textwidth]{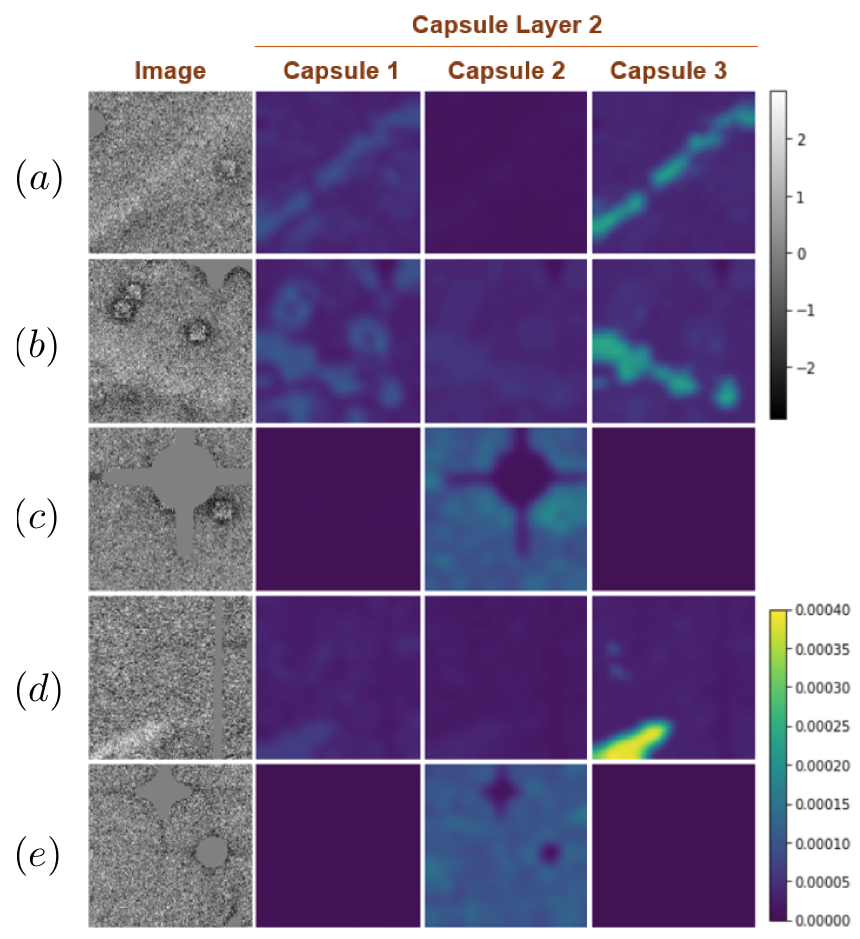}
    \caption{Routing path visualization to interpret the entities that the capsules in capsule layer 2 of model 8 detect. Sample images from test set. Sample test images are of size $200 \times 200$, and routing path visualizations of capsule layer 2 are of size $29 \times 29$. Images (a), (b), and (d) contain light echoes; (c) and (e) contain stars. All images show varying amounts of interstellar dust or artifacts. Illustration and caption credit: \citep{rpv}.}
    \label{fig:caps2_old}
\end{figure}

Figure \ref{fig:caps2_old} displays sample routing path visualizations of the capsules in capsule layer 2 for model 8 (Table \ref{table:model_caps}). The brighter the region in a routing path visualization the more that region of the image is routed to that particular capsule. Capsule 1 seems to be detecting a combination of light echoes and artifacts; capsule 2 seems to be detecting the diffraction pattern difference features around bright stars; capsule 3 seems to be detecting light echoes. Accordingly, the model may fail when given an image of a bright star because capsule 2 will become active, and the irrelevant information contained in capsule 2 will be routed to the light-echo-detecting-capsule. It was found that increasing the number of capsule types, $I$, and the number of dimensions per capsule, $D$, resulted in a cleaner routing path visualization. ALED-m and models 6, 8, and 9 had comparable accuracies on both the validation and test sets and were best performing among all capsule network and CNN architectures tested.

\begin{figure*}
    \centering
    \includegraphics[width=\textwidth]{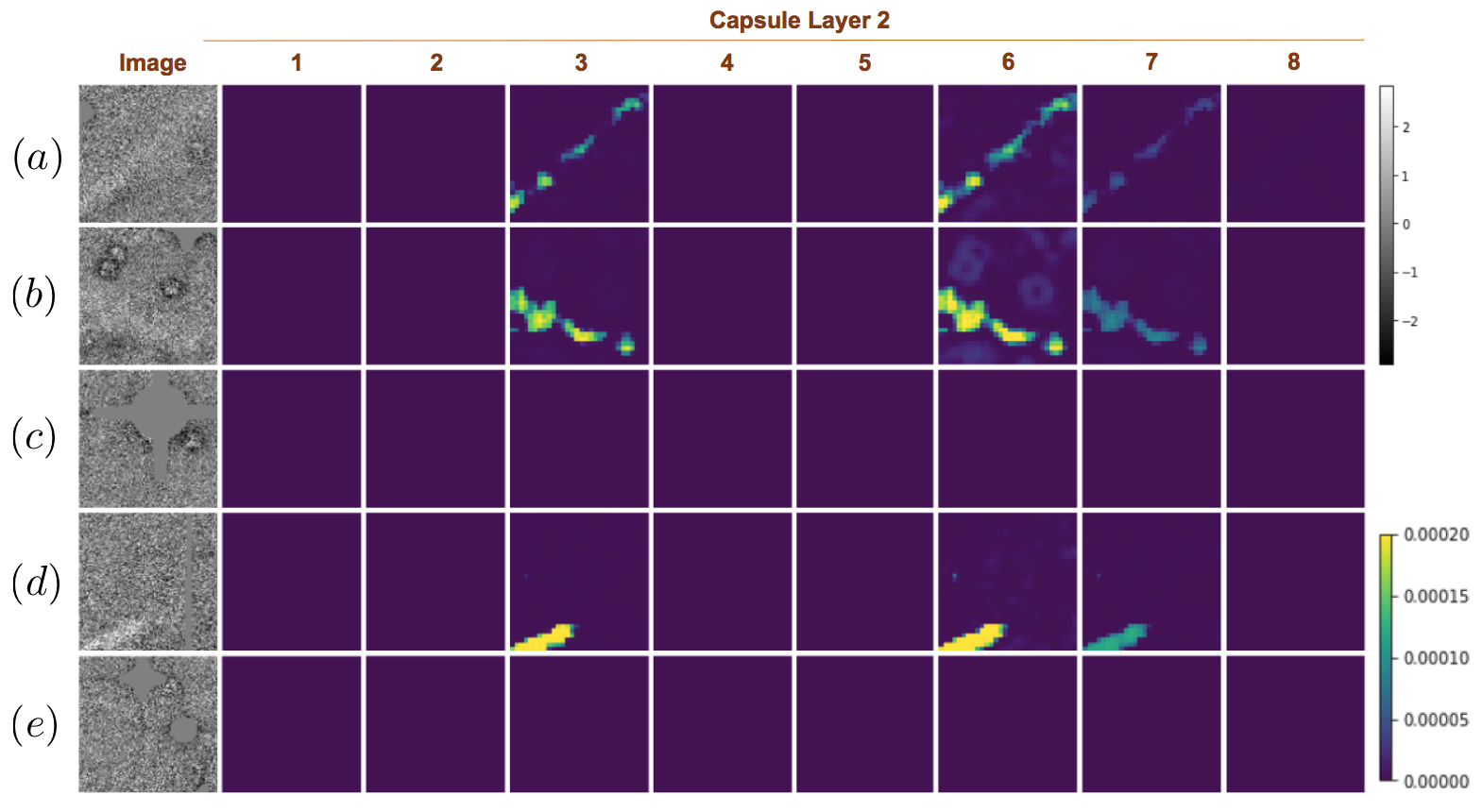}
    \caption{Routing path visualization to interpret the entities detected by the capsules in capsule layer 2 of ALED-m. Sample images from test set. On a set of sample test images of size $200\times200$. Routing path visualizations of capsule layer 2 are of size $29\times29$.}
    \label{fig:caps2}
\end{figure*}

The absence of artifacts in the routing path visualizations in Figure \ref{fig:caps2} in comparison to Figure \ref{fig:caps2_old} reveals that the capsules in capsule layer 2 of ALED-m are not nearly as sensitive to artifacts as model 8. This finding is reasonable because ALED-m is more complex than model 8, and thus, can learn a more appropriate strategy for classification. All capsules in capsule layer 2 seem to be detecting entities related to light echoes, which makes the model less susceptible to incorrectly classifying a non-light-echo image as a light echo image.

\begin{figure*}
\centering
\begin{subfigure}{.5\textwidth}
  \centering
  \includegraphics[width=0.52\linewidth]{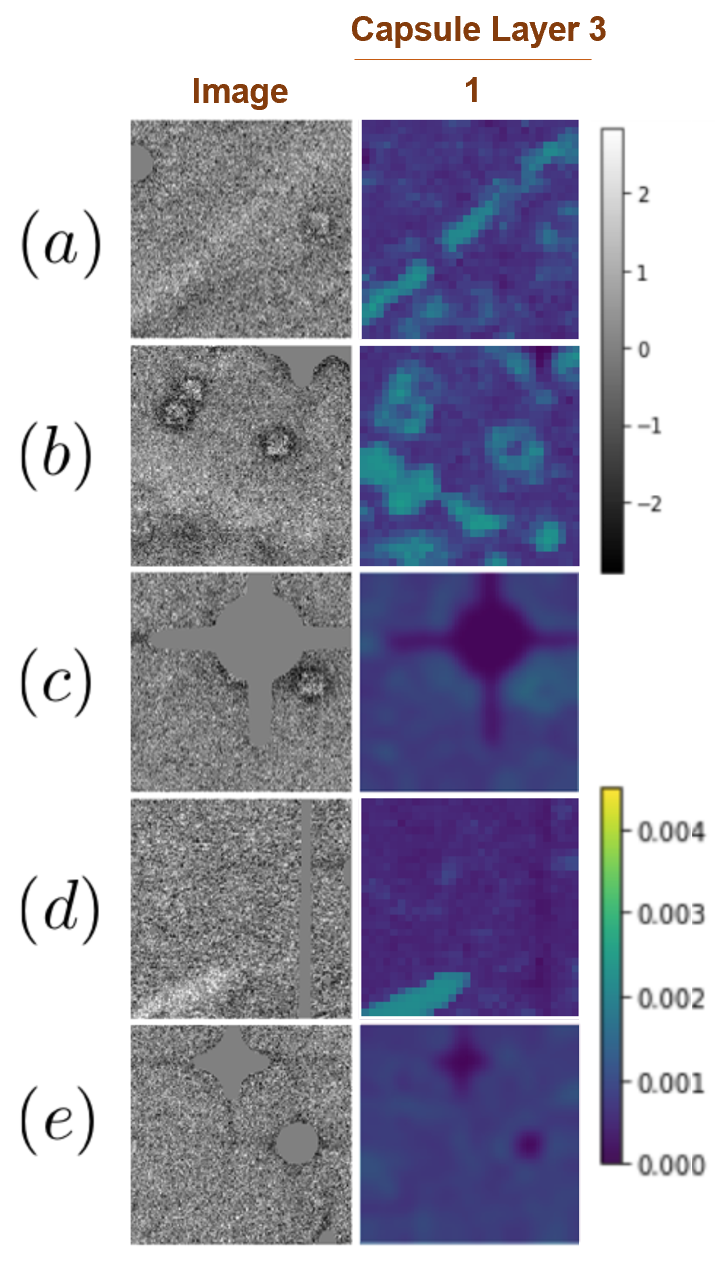}
  \caption{model 8}
  \label{fig:sub1}
\end{subfigure}%
\begin{subfigure}{.5\textwidth}
  \centering
  \includegraphics[width=0.5\linewidth]{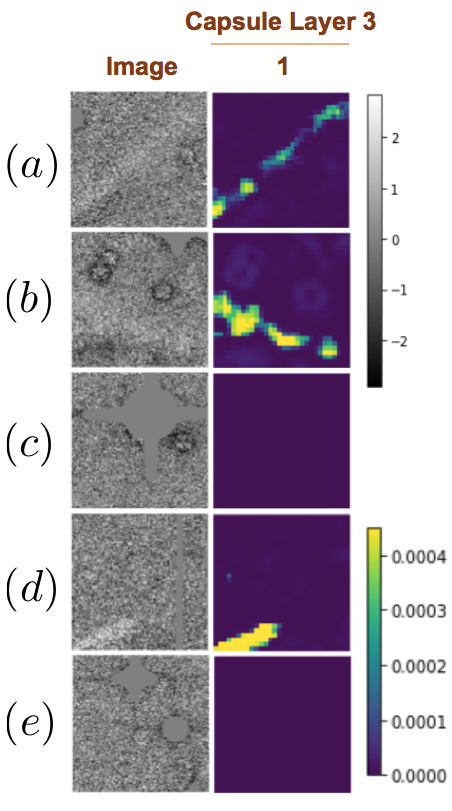}
  \caption{ALED-m}
  \label{fig:sub2}
\end{subfigure}
\caption{Routing path visualization to localize light echoes detected by the light-echo-detecting-capsule. Sample images from test set. Sample test images are of size $200\times200$, and routing path visualizations of capsule layer 3 are of size $29\times29$.}
    \label{fig:caps3}
\end{figure*}

From Figure \ref{fig:caps3}, it is evident that a routing path visualization of the light-echo-detecting capsule in ALED-m precisely localizes light echoes, if they are present in the image. On the contrary, a routing path visualization of the light-echo-detecting capsule in model 8 localizes other entities in addition to light echoes. It seems that a model with more weights can more cleanly localize light echoes in a routing path visualization. In short, model 8 does not always use the correct information in the image to predict the presence of a light echo because irrelevant regions of the image are routed to the light-echo-detecting capsule. Regardless, both model 8 and ALED-m were among the best performing models because of their low misclassification rates (model 8: 5/50, ALED-m: 4/50) and low margin loss (model 8: 0.037, ALED-m: 0.039). These models could better distinguish light echoes from artifacts compared to most other models. See Table \ref{table:model_confusion} of the Appendix for classification details for select models.

\renewcommand\thesubfigure{\roman{subfigure}}

\subsection{ALED Classification Results}

\begin{figure}
    \centering
    \includegraphics[width=0.45\textwidth]{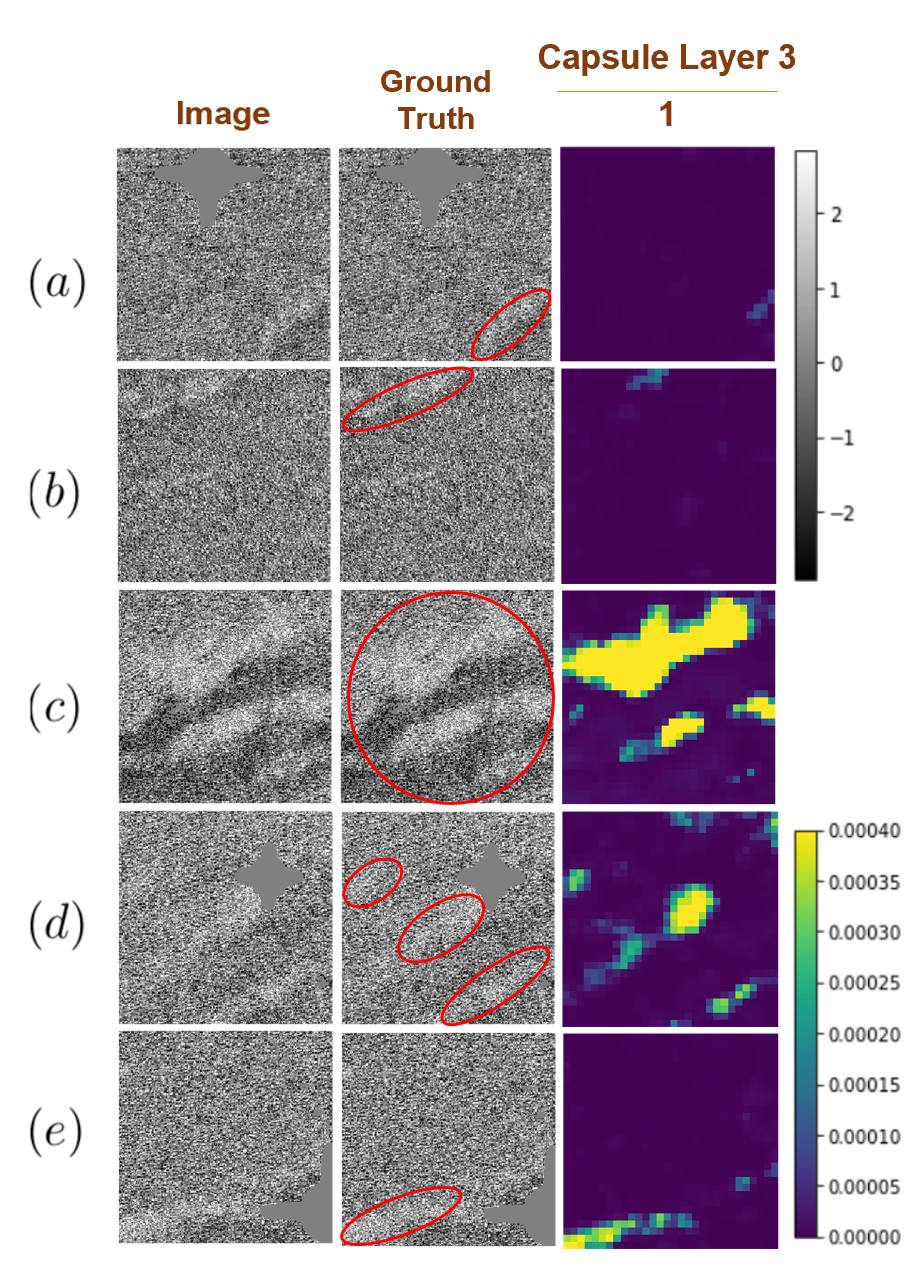}
    \caption{Sample of routing path visualizations of 5 light echos. (a)-(b) classified to "no light echo"; (c)-(e) classified to "light echo". Red ellipses localize light echoes.}
    \label{fig:brightness}
\end{figure}

ALED-m was found to have fewer false positives if classification was based on the routing path visualization pixel value, rather than on the length of the of the light echo-detecting capsule.  As such, if a routing path visualization contains at least one pixel with a value greater than 0.00042 then ALED classifies the corresponding difference image as a light echo candidate. However, this threshold can be changed to increase or decrease the pool of light echo candidates. ROC and precision-recall curves were generated by varying the classification threshold from 0 to 0.00070 (Figure \ref{fig:roc} of the Appendix), and then applying ALED-m to the entire set of 350 cropped CFHT images of size 200$\times$200. The confusion matrix corresponding to a threshold of 0.00042 shows that there are no false positives, so ALED-m is able to identify light echos with specificity of 1 (Table 4 and Figure \ref{fig:brightness} (c)-(e)). It was found that very faint and narrow light echoes may get falsely classified as non-light-echoes (Figure \ref{fig:brightness} (a)-(b)).

\begin{table}
\centering
\caption{The number of true positive (TP), true negative (TN), false positive (FP) and false negative (FN) classifications based on ALED-m, using pixel value threshold of 0.00042 for light echo classification.}
\scalebox{1}{
\begin{tabular}{ r r r r }
\toprule
\textbf{TP} & \textbf{TN} & \textbf{FP} & \textbf{FN} \\
\midrule
 44 & 175 & 0 & 131 \\

\bottomrule
\end{tabular}}
\label{table:model_confusion_aled}
\end{table}

All 13,000+ $2048 \times 4612$ difference images produced in 2011 by CFHT's MegaCam, including the 22 images that were used to train ALED-m, were also classified by ALED. When input images are larger than 200x200, ALED pads the input image so that it is completely divisible into 200 $\times$ 200 sub-images. Each sub-image is passed through ALED-m, and a corresponding routing path visualization is produced. The routing path visualizations are stitched together into a final routing path visualization.

\begin{figure}[H]
    \centering
    \includegraphics[width=.45\textwidth]{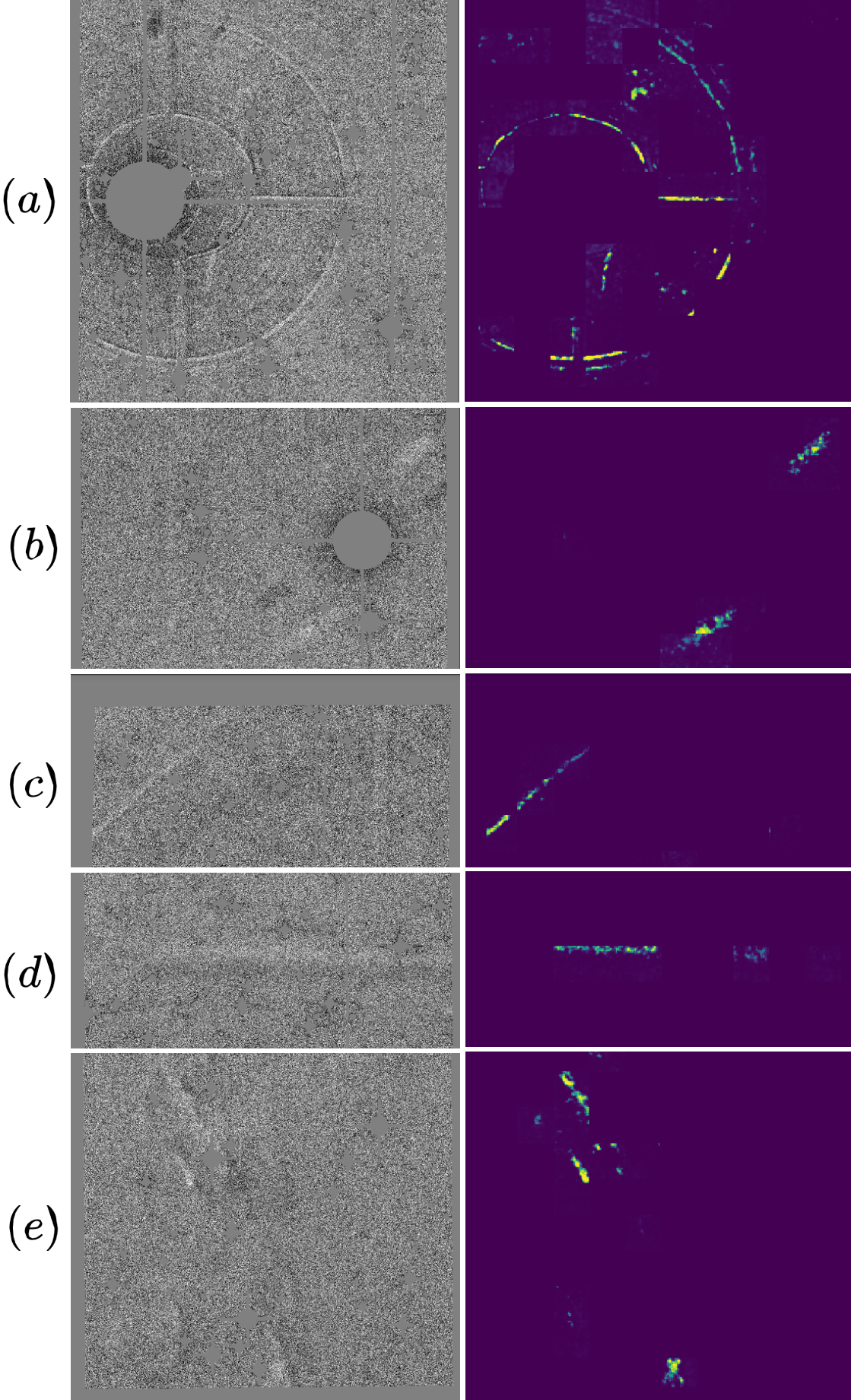}
    \caption{Difference images and their corresponding routing path visualizations. (a)-(d) sample artifacts that ALED classified as light echo candidates; (e) Cas A light echoes that ALED correctly classified.}
    \label{fig:arts}
\end{figure}

From the 13,000+ difference images, 1646 images were classified to be light echo candidates. The 1646 difference images were then manually inspected for validation -- a two-hour process, and one difference image was found to contain bona fide Cas A light echoes not noted by \citet{mcdonald}, see Figure \ref{fig:arts} (e). The significance of this find is that the ALED algorithm was able to identify a group of light echoes that were not in the training set.

About 40\% of the light echo candidates from the classification of all available difference images were due to bright stars, see Figure \ref{fig:arts} (a). There are always small pointing differences between the two images used to create a difference image. The diffraction patterns around bright stars are then slightly displaced and when differenced, show radially-displaced patterns that can be mistaken for very sharp, low apparent motion, light echo features. This mis-classification is reasonable because bright star artifacts are large, usually spanning a $1000 \times 1000$ pixel region of an image, and ALED only looks at local $200 \times 200$ regions of an image at a time. Fortunately, since the positions and apparent brightness of such stars is known, bright stars can be algorithmically filtered from the pool of light echo candidates. Figures \ref{fig:arts} (b)-(d) show the routing path visualizations of rejected light echo candidates, all of which exhibit light echo-like features. Apparently, several such features were contained in the training set but ALED correctly assessed them as unlikely light echo candidates.

\section{Summary} \label{section:paper2_summary}
We have presented ALED, a novel tool for automated classification of light echoes from differenced astronomical images. ALED uses capsule networks for classification and routing path visualization to localize regions of the image contributing to the classification.  Performance of ALED is competitive with CNNs, but requires far fewer training weights or training samples, and does not rely on either drop out or augmenting the training set with training image transpositions. For practitioners, the routing path visualizations facilitate quick identification of class entities within an image.

Model ALED-m is implemented in the Python package ALED and is trained on the CFHT data to predict presence of light echoes in astronomical images. ALED-m uses the correct regions of an image when predicting the presence of light echo and can precisely localize light echoes when they are present in an image.  The arduous task of classifying  13000+ images by manual inspection was drastically reduced by applying ALED-m to identify light echo candidates and then using manual inspection for confirmation only on the set of light echo candidates. Further, one novel light echo image was detected by ALED-m that had been missed by manual classification. 

Although ALED uses a capsule network to improve generalizability, future work involves testing ALED on astronomical images not taken from the CFHT MegaCam. The Python package ALED is publicly available at github.com/LightEchoDetection/ALED, though corresponding package documentation is also provided in Appendix C. To the authors' knowledge, this is the first automation of the laborious task of light echo detection.

\begin{acknowledgements}
 We are grateful to Dr. Armin Rest (STScI) for the preparation of the difference images from the CFHT MegaCam survey. This research was enabled in part by support provided by Compute Ontario (www.computeontario.ca) and Compute Canada (www.computecanada.ca). The images analyzed were based on observations obtained with MegaPrime/MegaCam, a joint project of CFHT and CEA/DAPNIA, at the Canada-France-Hawaii Telescope (CFHT) which is operated by the National Research Council (NRC) of Canada, the Institut National des Sciences de l’Univers of the Centre National de la Recherche Scientifique of France, and the University of Hawaii. This research was supported by the National Science and Engineering Research Council of Canada.
\end{acknowledgements}

\bibliographystyle{abbrvnat}
\bibliography{\jobname}

\begin{thebibliography}{15}
\providecommand{\natexlab}[1]{#1}
\providecommand{\url}[1]{\texttt{#1}}
\expandafter\ifx\csname urlstyle\endcsname\relax
  \providecommand{\doi}[1]{doi: #1}\else
  \providecommand{\doi}{doi: \begingroup \urlstyle{rm}\Url}\fi

\bibitem[Abadi et~al.(2016)Abadi, Agarwal, Barham, Brevdo, Chen, Citro,
  Corrado, Davis, Dean, Devin, et~al.]{abadi2016tensorflow}
M.~Abadi, A.~Agarwal, P.~Barham, E.~Brevdo, Z.~Chen, C.~Citro, G.~S. Corrado,
  A.~Davis, J.~Dean, M.~Devin, et~al.
\newblock Tensorflow: Large-scale machine learning on heterogeneous distributed
  systems.
\newblock \emph{arXiv preprint arXiv:1603.04467}, 2016.

\bibitem[Bhullar et~al.(2020)Bhullar, Ali, and Welch]{rpv}
A.~Bhullar, R.~A. Ali, and D.~L. Welch.
\newblock Interpreting capsule networks for image classification by routing
  path visualization.
\newblock In \emph{Submitted to IEEE Transactions on Pattern Analysis and
  Machine Intelligence}, 2020.

\bibitem[Goodfellow et~al.(2016)Goodfellow, Bengio, and
  Courville]{goodfellow2016deep}
I.~Goodfellow, Y.~Bengio, and A.~Courville.
\newblock \emph{Deep learning}, volume~1.
\newblock MIT press Cambridge, 2016.

\bibitem[Hinton et~al.(2012)Hinton, Srivastava, Krizhevsky, Sutskever, and
  Salakhutdinov]{dropout}
G.~E. Hinton, N.~Srivastava, A.~Krizhevsky, I.~Sutskever, and R.~R.
  Salakhutdinov.
\newblock Improving neural networks by preventing co-adaptation of feature
  detectors.
\newblock \emph{arXiv preprint arXiv:1207.0580}, 2012.

\bibitem[Kingma and Ba(2014)]{kingma2014adam}
D.~P. Kingma and J.~Ba.
\newblock Adam: A method for stochastic optimization.
\newblock \emph{arXiv preprint arXiv:1412.6980}, 2014.

\bibitem[LeCun et~al.(1989)LeCun, Boser, Denker, Henderson, Howard, Hubbard,
  and Jackel]{lecun1989backpropagation}
Y.~LeCun, B.~Boser, J.~S. Denker, D.~Henderson, R.~E. Howard, W.~Hubbard, and
  L.~D. Jackel.
\newblock Backpropagation applied to handwritten zip code recognition.
\newblock \emph{Neural computation}, 1\penalty0 (4):\penalty0 541--551, 1989.

\bibitem[LeCun et~al.(1998)LeCun, Bottou, Bengio, and
  Haffner]{lecun1998gradient}
Y.~LeCun, L.~Bottou, Y.~Bengio, and P.~Haffner.
\newblock Gradient-based learning applied to document recognition.
\newblock \emph{Proceedings of the IEEE}, 86\penalty0 (11):\penalty0
  2278--2324, 1998.

\bibitem[McDonald(2012)]{mcdonald}
B.~J. McDonald.
\newblock The search for supernova light echoes from the core-collapse
  supernovae of {AD} 1054 (crab) and {AD} 1181 (master’s thesis), 2012.
\newblock McMaster University.

\bibitem[Rest et~al.(2005)Rest, Stubbs, Becker, Miknaitis, Miceli, Covarrubias,
  Hawley, Smith, Suntzeff, Olsen, et~al.]{rest2005testing}
A.~Rest, C.~Stubbs, A.~C. Becker, G.~Miknaitis, A.~Miceli, R.~Covarrubias,
  S.~Hawley, R.~C. Smith, N.~B. Suntzeff, K.~Olsen, et~al.
\newblock Testing lmc microlensing scenarios: The discrimination power of the
  supermacho microlensing survey.
\newblock \emph{The Astrophysical Journal}, 634\penalty0 (2):\penalty0 1103,
  2005.

\bibitem[{Rest} et~al.(2015){Rest}, {Sinnott}, {Welch}, {Prieto}, {Bianco},
  {Matheson}, {Smith}, and {Suntzeff}]{2015ASPC..491..247R}
A.~{Rest}, B.~{Sinnott}, D.~L. {Welch}, J.~L. {Prieto}, F.~B. {Bianco},
  T.~{Matheson}, R.~C. {Smith}, and N.~B. {Suntzeff}.
\newblock {Light Echoes of Ancient Transients with the Blanco 4m Telescope}.
\newblock In S.~{Points} and A.~{Kunder}, editors, \emph{Fifty Years of Wide
  Field Studies in the Southern Hemisphere: Resolved Stellar Populations of the
  Galactic Bulge and Magellanic Clouds}, volume 491 of \emph{Astronomical
  Society of the Pacific Conference Series}, page 247, May 2015.

\bibitem[Sabour et~al.(2017)Sabour, Frosst, and Hinton]{sabour2017dynamic}
S.~Sabour, N.~Frosst, and G.~E. Hinton.
\newblock Dynamic routing between capsules.
\newblock In \emph{Advances in neural information processing systems}, pages
  3856--3866, 2017.

\bibitem[Schechter et~al.(1993)Schechter, Mateo, and Saha]{schechter1993dophot}
P.~L. Schechter, M.~Mateo, and A.~Saha.
\newblock Dophot, a ccd photometry program: Description and tests.
\newblock \emph{Publications of the Astronomical Society of the Pacific},
  105\penalty0 (693):\penalty0 1342, 1993.

\bibitem[Selvaraju et~al.(2017)Selvaraju, Cogswell, Das, Vedantam, Parikh, and
  Batra]{selvaraju2017grad}
R.~R. Selvaraju, M.~Cogswell, A.~Das, R.~Vedantam, D.~Parikh, and D.~Batra.
\newblock Grad-cam: Visual explanations from deep networks via gradient-based
  localization.
\newblock In \emph{Proceedings of the IEEE international conference on computer
  vision}, pages 618--626, 2017.

\bibitem[Shang et~al.(2012)Shang, Hu, Hu, Li, Li, Liu, Ma, Quinn, Sun, Wang,
  et~al.]{shang2012operation}
Z.~Shang, K.~Hu, Y.~Hu, J.~Li, J.~Li, Q.~Liu, B.~Ma, J.~L. Quinn, J.~Sun,
  L.~Wang, et~al.
\newblock Operation, control, and data system for antarctic survey telescope
  (ast3).
\newblock In \emph{Observatory Operations: Strategies, Processes, and Systems
  IV}, volume 8448, page 844826. International Society for Optics and
  Photonics, 2012.

\bibitem[Zagoruyko and Komodakis(2016)]{zagoruyko2016wide}
S.~Zagoruyko and N.~Komodakis.
\newblock Wide residual networks.
\newblock \emph{arXiv preprint arXiv:1605.07146}, 2016.

\end{thebibliography}


\newpage

\onecolumn
\appendix
\counterwithin{figure}{section}

\section{Reduction and Difference-Imaging Technique \citep{mcdonald}}

This appendix summarizes the pre-processing of astronomical photos to produce the differenced images on which ALED was trained. Difference-imaging involves the digital image subtraction of two or more epochs of the same field or viewpoint. McDonald (2012) developed an image pipeline that used reduced CFHT images and adapted a technique by \citet{rest2005testing} to perform the differencing. A brief description of the difference-imaging steps implemented by the image pipeline on the reduced images are described below:

\begin{enumerate}

\item \textbf{Deprojection} - Each image was resampled to the same geometry as the template image (earlier epoch image) so that images are photometrically aligned to facilitate subtraction per the SWarp software package \citep{shang2012operation}.

\item \textbf{Aperture Photometry} - The reduced and resampled images are photometrically calibrated using the DoPHOT photometry package to identify and measure sources \citep{schechter1993dophot}, and identify a photometric zero point for the image.

\item \textbf{Pixel Masking} - Saturated pixels are masked out since their true brightness values are unknown (i.e. for saturated stars, both the star and its spikes are masked).

\item \textbf{Image Subtraction} - Clean difference-images are produced by point-spread-function (PSF)-matching and then subtracting pixel values, using the Higher Order Transform of PSF and Template Subtraction (HOTPANTS) package. 

\end{enumerate}

The photonetric alignment and PSF-matching are the most difficult steps of this process. Indeed, the images to be subtracted are often taken under different conditions, including atmospheric transparency, atmospheric seeing, or exposure times, with each image consequently having different PSF. A convolution kernel that matches the PSFs of two astronomical images is found such that the output pixel is a weighted sum of the input pixels within a kernel of a certain size. Further, since the PSF of astronomical images are spatially varied, the kernel must be modelled as a spatially varying function \citep{mcdonald}.

\section{Extended Model Evaluation}

\begin{figure*}[h]
\centering
\begin{subfigure}{.5\textwidth}
  \centering
  \includegraphics[width=0.8\linewidth]{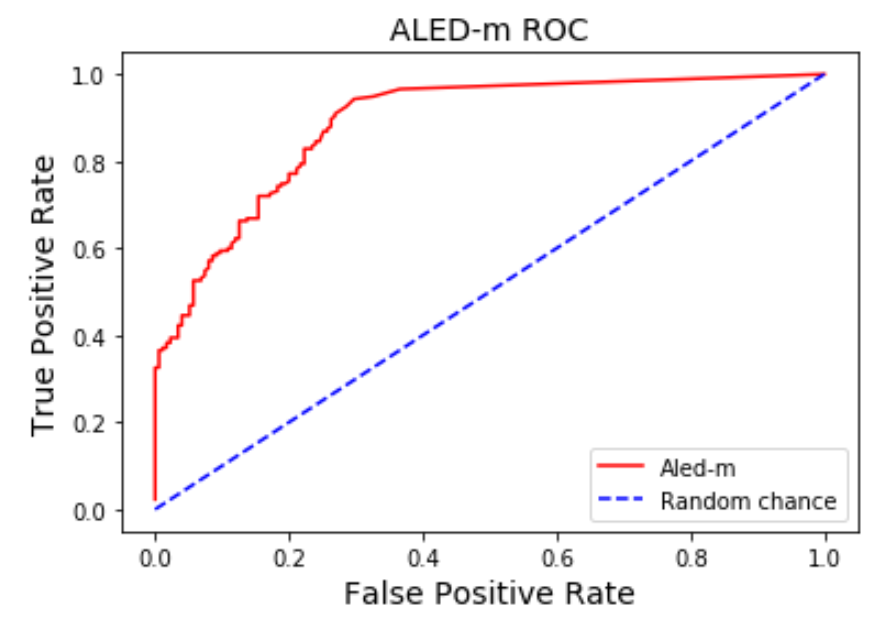}
  \caption{ROC curve}
  \label{figss:sub1}
\end{subfigure}%
\begin{subfigure}{.5\textwidth}
  \centering
  \includegraphics[width=0.8\linewidth]{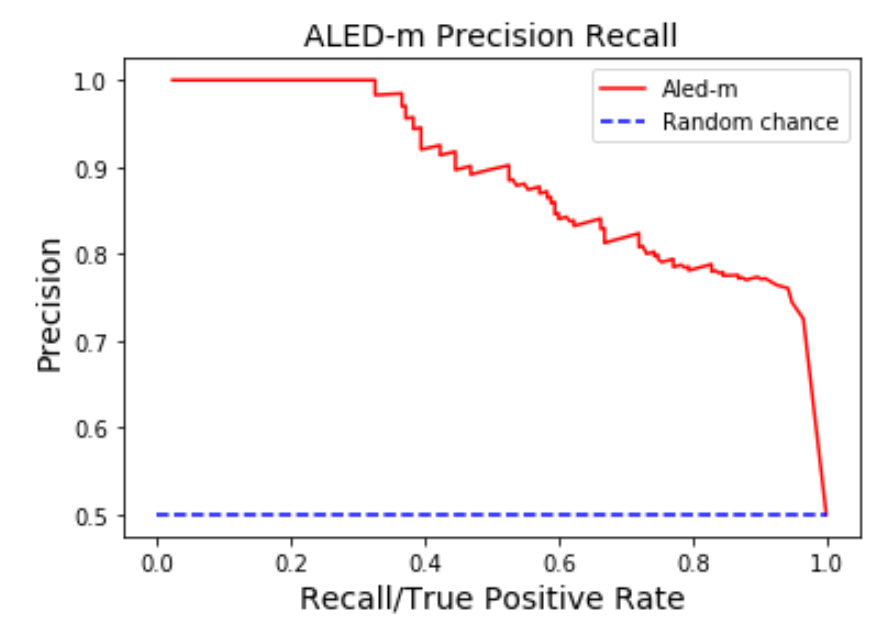}
  \caption{Precision-recall curve}
  \label{figss:sub2}
\end{subfigure}
\caption{The ROC and precision-recall curve of ALED-m as calculated by varying the threshold of 0.00042 from 0 to 0.00070.}
    \label{fig:roc}
\end{figure*}

Here we present further classification results. Table \ref{table:model_confusion} compares classification using different capsule and CNN architectures for classification threshold based on length of light echo-detecting capsule.  Lower model loss was typically associated with better ability to distinguish light echoes from light echo like artifacts. Figure \ref{fig:roc} plots performance curves for model ALED-m using classification threshold based on routing path visualization pixel value.

\begin{table*}[h]
\centering
\caption{Classification (TP: true positive, TN: true negative, FP: false positive, FN: false negative) of the test set of 50 images, for several models. For the capsule networks, an image was classified as containing a light echo if the length of the final capsule was greater than 0.5. For the CNNs, an image was classified as containing a light echo if the neuron corresponding to the light echo class was more active than the non-light-echo class neuron. The model loss is given by the margin loss for capsule networks, and cross-entropy loss for CNNs.}
\scalebox{0.9}{
\begin{tabular}{ c c r r r r r }
\toprule
 &  &  \multicolumn{4}{c}{\textbf{Classification}} \\
\cmidrule(lr){3-6} 
\textbf{Network Type} & \textbf{Model} & \textbf{TP} & \textbf{TN} & \textbf{FP} & \textbf{FN} & \textbf{Loss} \\
\midrule
Capsule & 2 & 23 & 20 & 5 & 2 & 0.041\\
& 3 & 25 & 0 & 25 & 0 & 0.202\\
& 4 & 23 & 21 & 4 & 2 & 0.046\\
& 5 & 23 & 18 & 7 & 2 & 0.056\\
& ALED-m & 23 & 23 & 2 & 2 & 0.039\\
& 6 & 23 & 23 & 2 & 2 & 0.041\\
& 7 & 23 & 20 & 5 & 2 & 0.047\\
& 8 & 24 & 21 & 4 & 1 & 0.037\\
& 9 & 23 & 22 & 3 & 2 & 0.213\\
& 10 & 23 & 19 & 6 & 2 & 0.045\\
\addlinespace

Convolutional & 12 & 21 & 22 & 3 & 4 & 0.371\\
& 13 & 21 & 22 & 3 & 4 & 0.459\\
& 14 & 17 & 16 & 9 & 8 & 0.644\\
& 15 & 12 & 18 & 7 & 13 & 0.721\\

\bottomrule
\end{tabular}}
\label{table:model_confusion}
\end{table*}

\section{Package Documentation - \protect\url{github.com/LightEchoDetection/ALED}}

This appendix provides details on how to use the ALED package and perform routing path visualization in Python.

The function \texttt{classify\_fits()} takes as input the path of a directory containing difference images, of any size, to be classified. Each difference image is cropped to many images of size $200 \times 200$, which are then classified and a corresponding routing path visualization is produced. The routing path visualizations are stitched together to form a final output routing path visualization that corresponds to the input image. The routing path visualization localizes the light echoes for the user. The output also includes a text file listing images that are good candidates for containing a light echo.

\subsection{Installation}
\begin{enumerate}
    \item Install Python 3.7 \\
    
    \item (Optional) create a virtual environment: \texttt{virtualenv aledpy}, and activate virtual environment: \texttt{source aledpy/bin/activate} \\
    
    \item  Install jupyter notebook: \texttt{pip install jupyterlab} \\
    
    \item Install dependencies via \texttt{pip install ...}
    
\begin{center}
Dependencies: \\
\texttt{astropy==3.0.5} \\
\texttt{matplotlib==3.0.2} \\
\texttt{numpy==1.16.3} \\
\texttt{opencv-python==3.4.4.19} \\
\texttt{pandas==0.24.1} \\
\texttt{scikit-image==0.14.2} \\
\texttt{scikit-learn==0.20.2} \\
\texttt{scipy==1.2.1} \\
\texttt{tensorflow-gpu==1.11.0} or \texttt{tensorflow==1.11.0} \\
\end{center}

Installing \texttt{tensorflow-gpu==1.11.0} isn't a straight forward \texttt{pip install tensorflow-gpu==1.11.0}, instead follow this tutorial https://www.tensorflow.org/install/gpu. \texttt{tensorflow==1.11.0} can be installed easily using \texttt{pip}, however, it is typically much slower than tensorflow-gpu because it only uses the cpu. \\

\item Run jupyter notebook: \texttt{jupyter notebook} \\

If you're remotely connected to the computer than port forward jupyter notebook onto your local computer via \texttt{local\_user@local\_host\$ ssh -N -f -L localhost:8888:localhost:8889 remote\_user@remote\_host} \\

6. To make sure everything is installed correctly, run \texttt{test.ipynb}

\end{enumerate}

\subsection{Description}
The \texttt{test.ipynb} file contains sample code to get you started. Call function \texttt{classify\_fits(snle\_image\_paths, snle\_names, start)} from file \texttt{model\_functions.py} to start the classification process. \texttt{snle\_image\_paths} is a Python list of the file paths of each differenced image to be classified (each image in .fits format). \texttt{snle\_names} is a Python list of the names of the images corresponding to the file paths (names can be arbitrary strings). \texttt{start} is an int that allows you to start the classification process where you left off, in case the process has to be terminated. \\

The input image will be cropped to multiple 200x200 sub-images (note that padding will be added to the input image so that it is completely divisible by 200x200). Each sub-image is passed through the network for classification, and a corresponding routing path visualization image is produced. The routing path visualization images are stitched together and saved as a .png in directory \texttt{asto\_package\_pics/}, along with the input image. \\

For each .fits image, a corresponding routing path visualization image will be saved to \texttt{astro\_package\_pics/}. In addition, a text file titled \texttt{snle\_candidates.txt} will be created. The text file contains the name of each .fits file, and 5 values called \texttt{Count1}, \texttt{Count2}, \texttt{Count3}, \texttt{Avg1}, \texttt{Avg2}, representing the liklihood of the image containing a light echo. From experience, if \texttt{Count1} is non-zero than the image should be considered a light echo candidate. \\

\begin{enumerate}
    \item \texttt{Count1}: A count of the number of pixels in the routing path visualization image that have a value greater than 0.00042. \\
    
    \item \texttt{Count2}: A count of the number of pixels in the routing path visualization image that have a value greater than 0.00037. \\
    
    \item \texttt{Count3}: A count of the number of pixels in the routing path visualization image that have a value greater than 0.00030. \\
    
    \item \texttt{Avg1}: Is the average length of the light-echo-detecting capsule for the \texttt{top\_n} sub-images with the largest length for the light-echo-detecting capsule. \\
    
    \item \texttt{Avg2}: Is the average length of the light-echo-detecting capsule for the \texttt{small\_n} sub-images with the largest length for the light-echo-detecting capsule.
\end{enumerate}

As default, \texttt{top\_n=45} and \texttt{small\_n=10}. \texttt{top\_n} and \texttt{small\_n} are arguments for \texttt{classify\_fits()} and can be changed via \texttt{classify\_fits(..., top\_n=45, small\_n=10)}. \\

\subsection{Test Package}

To check if the dependencies have been installed correctly open \texttt{test.ipynb} and run all cells. If successful, 3 files should be produced:
\begin{enumerate}
    \item \texttt{snle\_candidates.txt}: Will contain the following line \texttt{test.fits 375.000000 440.000000 562.000000 0.731215 0.880019}
    \item \texttt{astro\_package\_pics/rpv\_test.fits.png}: The routing path visualization image of \texttt{test.fits}
    \item \texttt{astro\_package\_pics/snle\_test.fits.png}: \texttt{test.fits} in .png format
\end{enumerate}

\subsection{Re-Train Model}
The user must have a dataset in order to re-train the model as one is not provided. The re-training script is titled \texttt{retrain\_script.ipynb} in this repository. We recommend that the dataset contain an equal number of light-echo and non-light-echo images to prevent any class imbalance issues that may arise. The script demonstrates how to train a new model with the same architecture as ALED-m, and starting from the weights learned by ALED-m.

\end{document}